\documentclass{article}
\usepackage{graphicx} % Required for inserting images
\usepackage{siunitx}
\usepackage{listings}
\usepackage{hyperref}
\usepackage{nomencl} % If needed
\makenomenclature
\usepackage{amsmath}
\usepackage{algorithm}
\usepackage{algpseudocode}
\usepackage{bm}
\usepackage{booktabs}
\usepackage[letterpaper, total={6in, 9in}]{geometry}
\usepackage{breqn}

\usepackage{multirow}
\usepackage{longtable}
\usepackage{booktabs}
\usepackage{svg}
\usepackage{wrapfig}

\title{Comparing the Performance of MC/DC's on-GPU Event-based Processing Methods in Multigroup and Continuous-energy Problems
\footnote{
This is an Accepted Manuscript of an article published by the American Nuclear Society in Proceedings of the International Conference on Mathematics and Computational Methods Applied to Nuclear Science and Engineering (M\&C 2025) on April 27-30 2025, available at: \url{https://doi.org/10.13182/MC25-47174}} \footnote{
Please cite as: B. Cuneo, J. P. Morgan, I. Variansyah, and K. E. Niemeyer. "Comparing the Performance of MC/DC's on-GPU Event-based Processing Methods in Multigroup and Continuous-energy Problems," in \textit{Proceedings of the International Conference on Mathematics and Computational Methods Applied to Nuclear Science and Engineering (M\&C 2025)}, pp.1944--1953. (2025). Denver, CO, USA. doi: \href{https://doi.org/10.13182/MC25-47174}{10.13182/MC25-47174}.
}}

\author{
    Braxton Cuneo\textsuperscript{1,2}\footnote{Contact: bcuneo@seattleu.edu}
    \and 
    Joanna Piper Morgan\textsuperscript{1,3}
    \and
    Ilham Variansyah\textsuperscript{1,4}
    \and
    Kyle E. Niemeyer\textsuperscript{1,3}
}

\date{%
    \small{
    \textsuperscript{1}Center for Exascale Monte Carlo Neutron Transport;  \\
    \textsuperscript{2}Dept. of Computer Science, Seattle University, Seattle, WA;  \\
    \textsuperscript{3}School of Mechanical, Industrial, and Manufacturing Engineering, Oregon State University, Corvallis, OR;\\
    \textsuperscript{4}School of Nuclear Science and Engineering, Oregon State University, Corvallis, OR
    }
}

\begin{document}

\maketitle

\begin{abstract}
    Monte Carlo / Dynamic Code (MC/DC) is a portable Monte Carlo neutron transport package for rapid numerical methods exploration in heterogeneous and HPC contexts, developed under the auspices of the Center for Exascale Monte Carlo Neutron Transport (CEMeNT).
    To support execution on GPUs, MC/DC delegates resource and execution management to Harmonize (another CEMeNT software project). 
    In this paper, we describe and compare the performance of the two methods that Harmonize currently provides: a stack-based method and a distributed, asynchronous method.
    As part of this investigation, we analyze the performance of both methods under the 3D C5G7 \textit{k}-eigenvalue benchmark problem and a continuous-energy infinite pin cell problem, as run across 4 NVIDIA Tesla V100s.
    We find that the asynchronous method exhibits stronger early scaling compared to the stack-based method in the 3D C5G7 benchmark.
    We also found that the asynchronous method exhibits mixed performance relative to the stack-based method in the continuous-energy problem, depending upon tally resolution, particle count, and transport loop decomposition.
\end{abstract}

\section{Introduction}\label{sec:1}

Monte Carlo / Dynamic Code (MC/DC) \cite{transport_cement_mcdc_2024} is a portable Monte Carlo neutron transport package, developed under the auspices of the Center for Exascale Monte Carlo Neutron Transport (CEMeNT).
MC/DC applies a novel combination of approaches to the problem of Monte Carlo particle transport to facilitate rapid numerical methods exploration in heterogeneous and HPC contexts \cite{morgan_monte_2024}. It is written in Python, supports both interpreted and JIT-compiled execution across a variety of CPU architectures via Numba \cite{variansyah_mc23_mcdc,lam_numba_2015}, and provides domain replication and decomposition through MPI.

In addition to these design choices, MC/DC implements its transport logic in a single repository covering both CPU and GPU execution.
Harmonize \cite{harmonize}, another CEMeNT project, enables this by abstracting GPU execution through an asynchronous runtime interface.
Through this interface, MC/DC defines its transport logic in Python functions, which are then transformed into GPU-compatible code via Numba.
In this manner, the logistics of evaluation are separated from the primary concerns of MC/DC developers.
This is in an effort to enable more rapid numerical methods development for Monte Carlo neutron transport on Exascale supercomputers.

Harmonize provides multiple evaluation schemes, which may be used interchangeably, so long as they are compatible with the available GPU architecture.
At the time of writing, two algorithms are available: a conventional event-based processing scheme that manages events through stacks and an experimental ``asynchronous'' scheme, which performs distributed organization and load balancing between warps without inter-block synchronization.

In this work, we outline the event-based approach's underlying theory and how it relates to the broader concept of \textbf{thread-data remapping (TDR)} \cite{TDR}.
We then discuss how Harmonize abstracts thread-data remapping through its runtime interface, how its algorithms implement that interface, and how those implementations relate to the evaluation of MC/DC's transport logic.
After this, we investigate the relative performance of MC/DC when evaluated with Harmonize's algorithms; specifically, we gather performance data for both under the 3D C5G7 \textit{k}-eigenvalue benchmark problem and a continuous-energy infinite pin cell problem, as run on a compute node with 4 NVIDIA Tesla V100s.
For the continuous-energy problem, we additionally investigate the effects of alternative event decompositions of the particle transport problem.

\section{Background: Thread-Data Remapping}\label{sec:2}

Architecturally, GPUs represent a trade-off between control hardware and processing hardware.
Unlike CPUs, which conventionally split threads of execution across independent control hardware, GPUs group multiple threads of execution into sets, called \textit{warps}, which share a program counter \cite{gpu_hardware}.
Threads within the same warp are organized by the same control hardware and are executed together as vector operations.
This alternative scheme is known as single-instruction multiple-thread (SIMT) processing, and it allows GPUs to provide more processing hardware without a proportional increase in control hardware. This provides higher processing throughput relative to die area and power consumed \cite{gpu_hardware}.

However, this shift to vector processing results in limitations on which algorithms may be efficiently evaluated.
Since threads in the same warp are coordinated by the same program counter, the threads within a warp can only ever be ``located'' at the same instruction in a program, and so must perform the same instruction as their neighbors.
To allow for branching while sharing a program counter, warps with threads taking different paths will evaluate each path serially, masking the side effects of the corresponding instructions for threads which don't belong to that path.
This phenomenon, where threads in the same warp take different paths through a program, is called \textbf{divergence} \cite{TDR}.
The serialized branching caused by divergence can result in adverse and unintuitive effects, namely that the runtime of a GPU thread can be affected by its neighbors, even if they are performing completely independent calculations.

Developers optimizing for performance on GPUs can deal with divergence in one of three ways:

\begin{itemize}
    \item Accept the overhead introduced by divergence;
    \item Alter the program's underlying set of tasks to remove the possibility of divergent paths; or
    \item Assign the program's tasks such that threads within the same warp take similar paths.
\end{itemize}

In cases where divergence is severe but the benefits of expensive branches or loop iterations outweigh their cost, the third technique may be employed. This method---known as the \textbf{event-based} approach in particle transport, and more generically as \textbf{TDR} \cite{TDR} in computer science---is the central concern of this work.

In principle, if a program can predict what processing needs to occur for each element in a set of data, elements can be grouped to optimize the similarity of branching within each warp.
This represents a trade-off, exchanging the cost of task analysis and assignment for reduced divergence in the assigned tasks and, hence, faster processing.
As tasks are assigned to threads, each task's data must also be assigned to that task's thread, establishing a thread-data mapping.

If a program cannot predict all processing ahead of time, the program may process each piece of data until the next task is known.
Once this new set of predicted tasks has been established, those tasks may be reassigned across warps to reduce divergence.
By repeating this process, a program may iteratively ``discover'' how data will be processed and perform thread-data remapping as divergence arises.

TDR requires programmers to subdivide a program into distinct tasks, with transitions between tasks representing points where data may be remapped.
Ideally, the program should be split into tasks that do not diverge much locally, with the most expensive points of divergence occurring on task boundaries.
In Monte Carlo particle transport, these tasks represent the ``events'' of the event-based approach.
By extension, different thread data remapping algorithms represent different variants of event-based processing.

\section{Thread-Data Remapping in MC/DC and Harmonize} \label{sec:2.2}

Internally, MC/DC delegates TDR to Harmonize, a hardware abstraction framework for GPUs also developed by CEMeNT.
Harmonize abstracts task execution on GPUs through an an asynchronous runtime interface, with execution facilities exposed through said interface.

A core capability exposed through this interface is the creation of independent logical threads on the GPU, which can evaluate arbitrary functions.
This evaluation is asynchronous: thread creation immediately returns once an object representing that thread has been registered with the runtime.
In the context of asynchronous runtimes, these objects are referred to as \textbf{promises}.
Any combination of input types can be encoded as a promise, but MC/DC only uses promises to represent particles, so they will be referred to as particles here.

Harmonize's asynchronous call interface leverages the fact that a TDR system is essentially a specialized asynchronous runtime; both maintain a set of tasks that must be performed. This set of tasks can be extended at any time by the threads it executes, and parent threads do not have any guarantees of when or on what processor its child thread is executed.

With this abstract interface, Harmonize currently provides two different runtime types (TDR methods).
The first of these runtime types is analogous to the conventional event-based execution method originally proposed by William Martin and Forrest Brown \cite{brown_stack}, which uses a collection of stacks to organize particle data destined for different event types. The second type of runtime applies a novel approach, which load balances and remaps data in an asynchronous, distributed manner. More detailed coverage of this system's design, as well as its performance characteristics, are delineated by Cuneo and Bailey \cite{brax2023}.

Since both algorithms could arguably be considered variants of the event-based approach, we will refer to the first method as \textbf{stack-based remapping} and the second method as \textbf{asynchronous remapping}.

Both remapping types benefit from an additional feature: Harmonize has a built-in set of functions that are automatically called in response to certain conditions. Two of these functions, \texttt{\textbf{initialize}} and \texttt{\textbf{finalize}}, serve to set up and close out any custom information that is tracked either per block or per hardware thread. More crucially, the built-in function \texttt{\textbf{make\_work}} is called by a warp when it is starved of work.

This \texttt{\textbf{make\_work}} function serves a similar role to the ``source event'' proposed by Hamilton et al. \cite{hamilton}, but is instead codified as an official entry point for work.
This allows users to provide a single definition of \texttt{\textbf{make\_work}}, which handles both initial work creation and work creation for load balancing.
Through the use of atomic counters and a hash-based RNG scheme \cite{rng}, MC/DC generates particles on-demand across blocks via \texttt{\textbf{make\_work}}, and creates new tasks for them through the asynchronous call interface.

\subsection{Stack-based Remapping} \label{sec:2.2.1}

In MC/DC, stack-based remapping performs binning of particles by event type, where each event type has one global set of input and output buffers.
These buffers operate as pop-only and push-only stacks, where the input buffer provides particles that have not yet been processed, and the output buffer stores particles that have already been processed.
Insertion into and removal from these buffers is mediated by atomic operations tracking the size of each stack, allowing threads within the same kernel launch to concurrently claim or deposit work on any buffer.
These push and pop operations are performed en masse by warps, with the first thread in the warp performing an atomic addition on the corresponding integer value to determine the offset of the first element being popped or pushed and the remaining threads reading/writing values relative to that first element. To reduce access times, particles are stored in the local memory of their corresponding thread when they are being processed for their respective event.

\iffalse
\begin{figure}[h]
    \centering
    \includegraphics[width=0.75 \textwidth]{img/stack_style.pdf}
    \caption{A diagram illustrating a simplified representation of how MC/DC's stack-style remapping organizes and marshals particle data.}
    \label{fig:enter-label}
\end{figure}
\fi

To avoid buffer overruns due to the overproduction of work, blocks only call \texttt{\textbf{make\_work}} if all particles across all input buffers have been processed.
After this, \texttt{\textbf{make\_work}} is called until the most full output buffer exceeds a threshold value, at which point the current kernel launch concludes.
This threshold value is instituted because a near-capacity input buffer can easily cause buffer overruns in cases where the corresponding particles produce descendants beyond their replacement rate.

\subsection{Asynchronous Remapping} \label{sec:2.2.2}

In contrast to stack-based remapping, asynchronous remapping uses shared memory to perform its remapping operations, reserving global memory as a method of last resort for particle data storage.
In asynchronous remapping, particles are organized through data structures called \textbf{work links}.
Each work link contains a warp-sized array of particles -- with one particle for every thread in the architecture's warp.
At any given time, a work link's array can only contain particles destined for the same event type, meaning they will be processed by the same async transport function from MC/DC.
Work links also include metadata about their respective array, including a unique identifier for the associated event type, the number of elements in the array that are filled with real particle data, and a ``next'' pointer.
With the event ID and particle count metadata, the runtime is capable of determining which partially filled work links may be safely coalesced while maintaining event homogeneity.
The ``next'' pointer included with each work link allows these arrays to be grouped into linked lists.
These linked lists serve to organize the arrays into collections and also allow for a variety of lock-free operations via atomics.

A Harmonize runtime manages work links through arena allocation \cite{arena}.
All work links in a Harmonize runtime's global memory are stored in a single array, and allocation requests for work links in global memory are served out of that array.
Likewise, all work links in each warp's shared memory are stored in a single array, and allocation requests for work links in shared memory are served out of the array owned by the executing warp.
Each of these arrays is managed by separate allocators, which track the allocation states of the contained nodes and organize them through linked lists.

The work link allocator type used per warp---referred to as a stash---serves not only to track which work links are in use, but also organizes work links to track the state of execution and to aid in the efficient packing of particles into work links.
Each warp's stash maintains a list of free links that contain no particles, a list of full links that have no unused elements, an index tracking the current work link, and a table mapping event IDs to partially filled work links in the stash.

Through this table of partially filled work links, a stash is capable of maintaining optimal packing across all arrays.
If a particle destined for a certain event is to be stored in the stash, and a partially filled work link of same-event particles is present, that work link is used to store the particle.
Once a work link is completely filled, it is pushed onto the list of full links, which is used by the warp as a supply of closely-packed low-divergence particle operations.
Once a work link's particles have been processed, that link is pushed onto the free list, allowing it to be recycled into storage for particles subsequently generated during processing.

Ideally, all particle data would be stored within the stash for its entire life span, since shared memory provides much faster access times compared to global memory.
However, in cases where particle production exceeds stash capacity, some of the particle data present in the stash is spilled to work links in global memory.
Likewise, in cases of severe starvation where \texttt{\textbf{make\_work}} can no longer produce particles, a warp may transfer particles from global memory into its stash for processing.

\begin{figure}
    \centering
    \includegraphics[width=.6\textwidth]{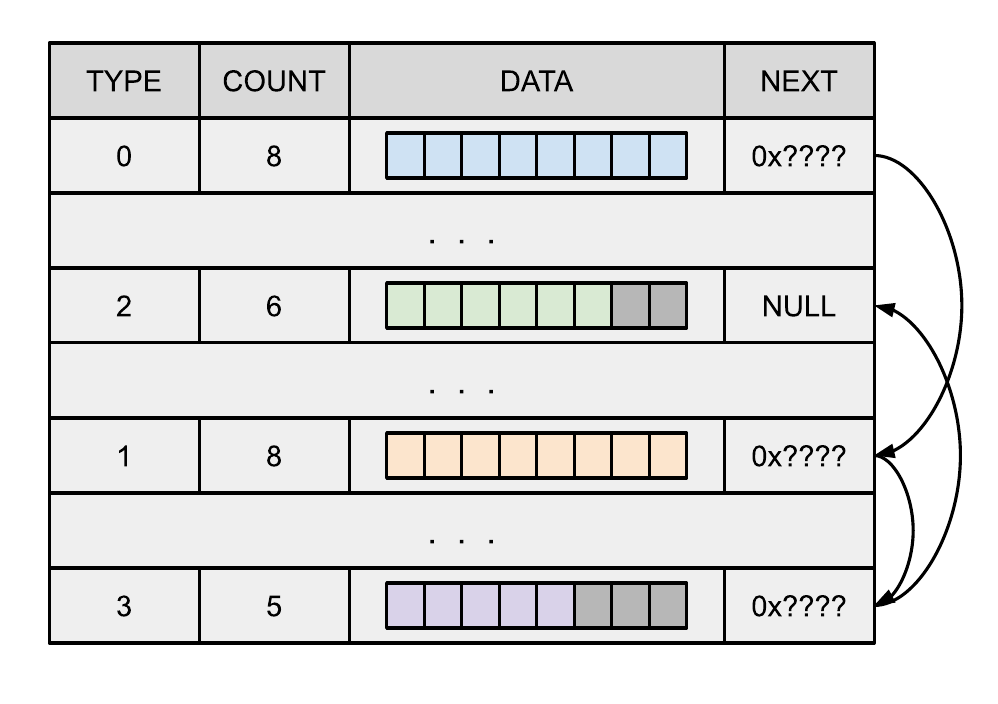}
    \caption{A diagram illustrating how work links connect into lists, reproduced with permission from Ref. \cite{brax2023}.}
    \label{fig:enter-label}
\end{figure}

Traditionally, if data needs to be exchanged between blocks, some form of synchronization is used to ensure the safe communication of data.
However, synchronizing between blocks can be rather expensive, requiring one block to wait until another block reaches the point of synchronization.
To avoid synchronization, work transfer between shared and global memory is accomplished in a lock-free fashion.

\begin{figure}
    \centering
    \includegraphics[width=.8\textwidth]{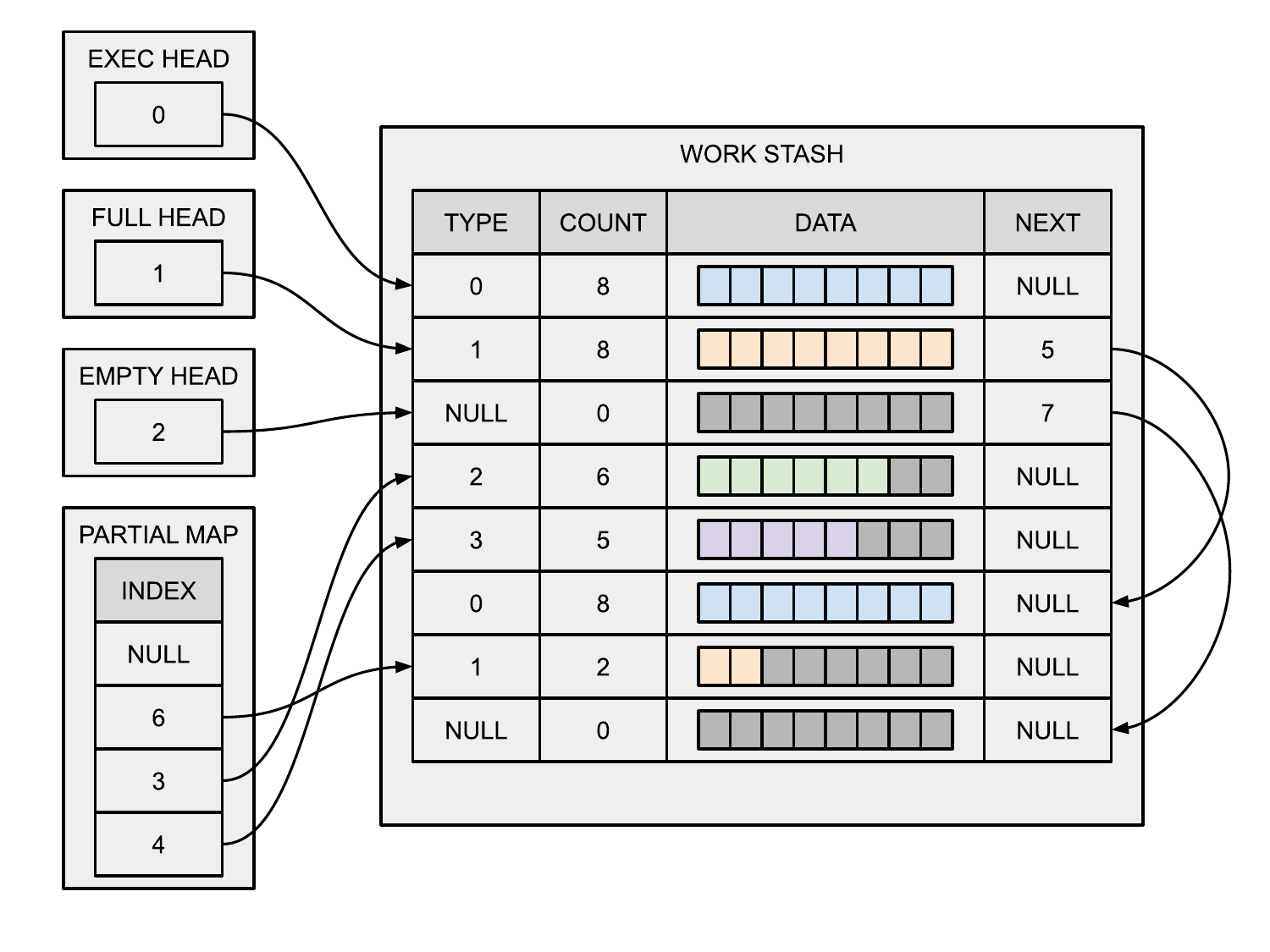}
    \caption{A diagram illustrating an example stash state, reproduced with permission from Ref. \cite{brax2023}.}
    \label{fig:enter-label}
\end{figure}

The global work link allocator divides its work links across a set of queues, each consisting of a pair of indices locating the head and tail of a work link list. One set of queues stores empty links, whereas the other stores non-empty links.
These queues pack both indices into a single integer and so can be atomically swapped with other queues as well as null-indexed empty queues.
By swapping a queue out with an empty queue, a warp may claim a queue temporarily, allowing it to add or remove links from that queue and return it through another atomic swap.
The process of transferring particles to global memory involves: (1) claiming both a queue of empty links and a queue of non-empty links, (2) transferring particles to the work links from the empty-link queue, (3) transferring these newly filled links to the non-empty-link queue, and (4) finally depositing both queues at their original location.
By reversing this process, a warp may likewise transfer particles from this reservoir back into shared memory.

\begin{figure}
    \centering
    \includegraphics[width=\textwidth]{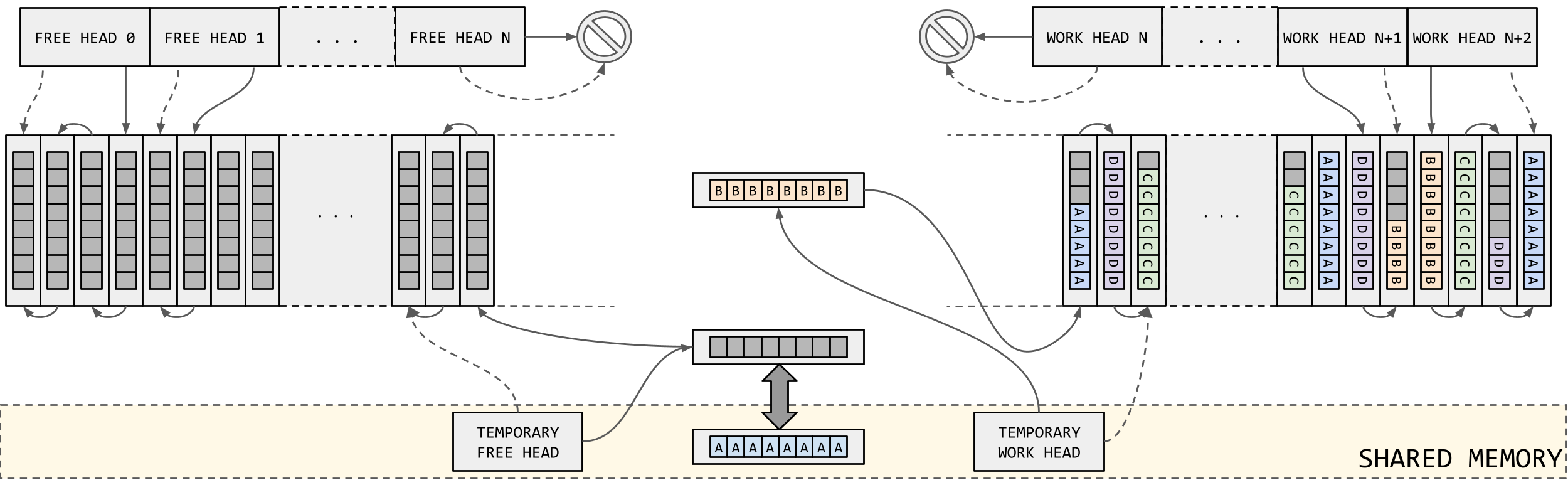}
    \caption{A diagram illustrating how work is transferred between shared memory and global memory under the asynchronous remapping method, reproduced with permission from Ref. \cite{brax2023}.}
    \label{fig:enter-label}
\end{figure}

\section{GPU Performance Analysis} \label{sec:4}

To investigate the relative performance of asynchronous and stack-based remapping, we gather runtime data for two benchmark problems.

The first of these benchmarks is the fully 3D C5G7 benchmark problem with multigroup cross-sections and geometries as defined by Hou et al. \cite{hou2017}.
With this problem, we run \textit{k}-eigenvalue simulations, using splitting roulette population control to bound storage requirements for particle data.
These simulations were run over 24, 48, 100, 200, and 400 cycles, with the first 25\% of cycles inactive and the remaining cycles active.
For each of these cycle counts, we ran simulations across a range of particle histories per cycle.

The second benchmark problem we consider models an infinite uranium dioxide (UO$_{\text{2}}$) fuel pin at 2.4\% enrichment surrounded by a homogeneous mixture of water and boron, with a uniform 14 MeV source.
This problem is modeled with continuous energy, and its full definition is provided in \texttt{examples/fixed\_source/} \texttt{inf\_pin\_ce/input.py} in v0.11.1 of MC/DC's
repository \cite{transport_cement_mcdc_2024}.
As with the C5G7 benchmark, this second benchmark problem was run across a range of source particle counts.
For each of these particle counts, this problem was run across a range of temporal tally resolutions.
This range begins with a mesh width of one along the time dimension, progressing by powers of two until a width of 128 along the time dimension.

To further refine our comparisons for this second benchmark, we employ two different task decomposition strategies.
The first is the approach currently employed by MC/DC in GPU execution, which consists of a single event type, called \textbf{step}.
The step event represents a single iteration of the typical history-based execution loop, otherwise known as a single segment of transport.
This single event, combined with the \texttt{\textbf{make\_work}} callback function, is quite similar to the history-length-truncation method investigated by Hamilton et al. \cite{hamilton} if the truncation length was set to one.

\begin{figure}
    \centering
    \begin{minipage}{0.153\textwidth}
    \frame{\includegraphics[width=\textwidth]{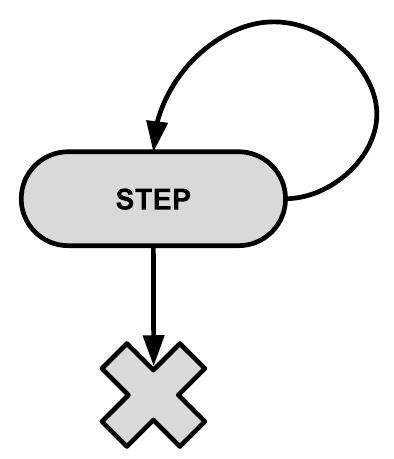}}
    \end{minipage}
    \begin{minipage}{0.575\textwidth}
    \frame{\includegraphics[width=\textwidth]{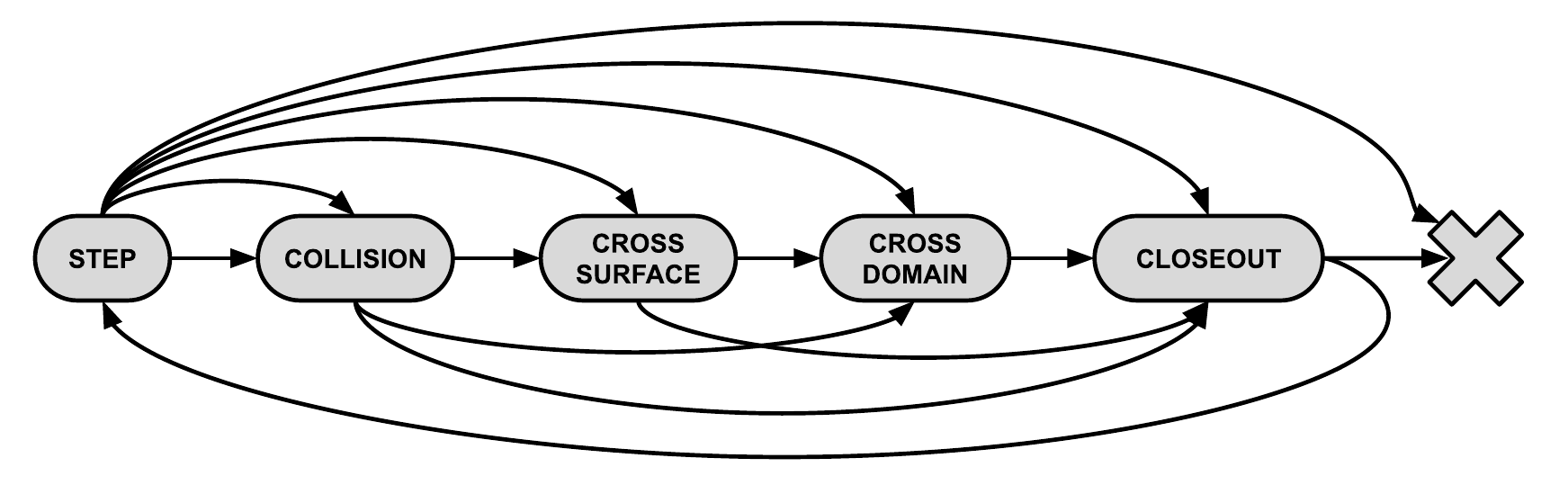}}
    \end{minipage}
    \caption{A diagram of potential paths of flow control under the tested monolithic task scheme (left) and the decomposed task scheme (right).}
    \label{fig:enter-label}
\end{figure}

The second decomposition strategy investigated in this work subdivided the history loop into five separate events. 
Under this scheme, the step event corresponds specifically to the process of moving a particle to the location of the next event and performing necessary tallies along that path.
Additional events correspond to collision events, surface crossing events, domain crossing events, and the ``closeout'' portion of the MC/DC particle transport loop, which handles weights and weight windows for the particle's next loop iteration.

We will refer to the first as the \textbf{monolithic task scheme} and the second as the \textbf{decomposed task scheme}.

\subsection{Platform Used for Data Collection}

We ran these problems on the Lassen machine available at Lawrence Livermore National Laboratory (LLNL).
Lassen has four NVIDIA Tesla V100s and two IBM Power 9 CPUs per node.

\section{Results and Analysis} \label{sec:5}

% mulitgroup problem
\subsection{C5G7 Benchmark Performance Comparison} \label{sec:5.1}

\begin{figure}
    \centering
    \includegraphics[width=\textwidth]{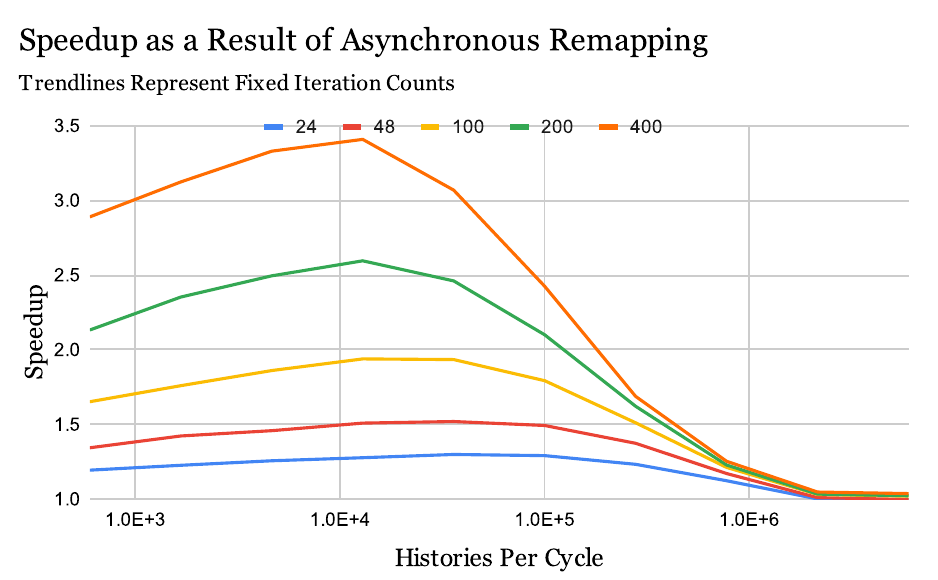}
    \caption{A graph depicting the speedup conferred by asynchronous remapping on the previously described 3D C5G7 \textit{k}-eigenvalue problem for 24, 48, 100, 200, and 400 iterations, respectively.}
    \label{fig:3D_C5G7}
\end{figure}

When evaluating the 3D C5G7 benchmark problem, both algorithms converge to the same upper-performance limit, approximately $6.9 \times 10^5$ histories per second (Fig \ref{fig:3D_C5G7}).
However, the asynchronous remapping method approaches this upper-performance limit more rapidly,
resulting in significant speedups relative to the stack-based remapping method.  
This speedup increases significantly as the number of iterations increases, with performance tripling in some cases. 

This initial performance increase implies that asynchronous remapping has a better strong scaling efficiency relative to stack-based remapping.
We posit that this stronger early scaling is due to the locality provided by the stash data structures used in asynchronous remapping.
As shown in prior research \cite{brax2023}, problems that do not exhibit rapid growth in simulated histories use less global memory for particle storage with asynchronous remapping versus stack-based remapping.
Since this problem is run without implicit capture, a relatively short time to census, and a relatively low \textit{k}-effective, stashes of the asynchronous runtime would likely not frequently spill particle data.

The shared upper limit upon performance between asynchronous and stack-based remapping implies that some other factor is limiting performance.
The most likely candidates for this limit are the actions prescribed by the transport logic.
For example, if MC/DC's transport logic performs many loads from global memory for a given problem, the latency of these loads would dominate the performance of the simulation, and hence offset benefits provided by asynchronous remapping.
Establishing the source of this performance limit will require a more thorough investigation, through methods like kernel profiling.

% continious energy problem
\subsection{Continuous Energy} \label{sec:5.2}

\begin{figure}
    \begin{minipage}{0.49\textwidth}
    \includegraphics[width=\textwidth]{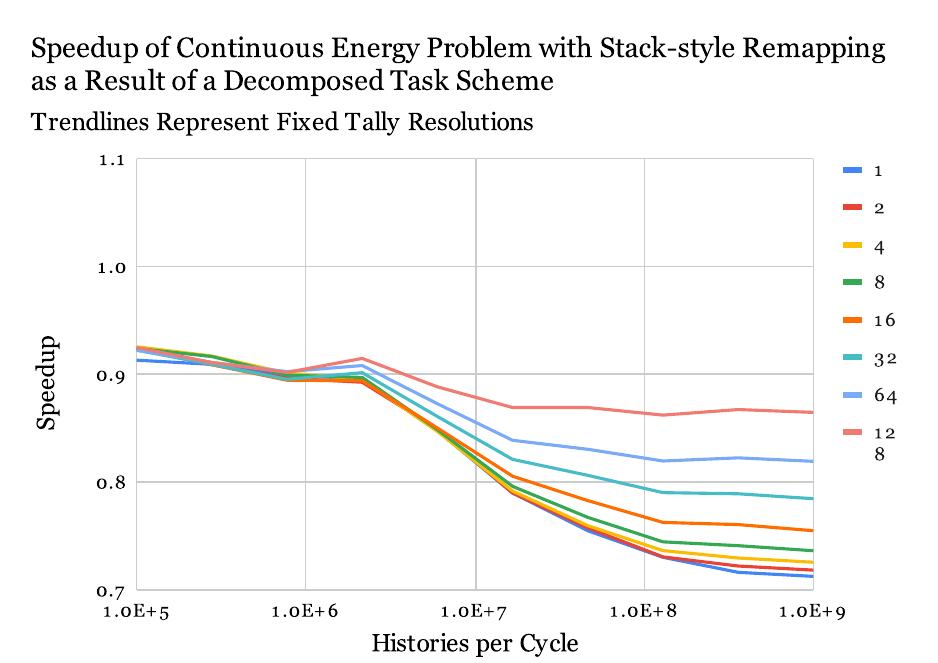}
    \end{minipage}
    \hfill
    \begin{minipage}{0.49\textwidth}
    \includegraphics[width=\textwidth]{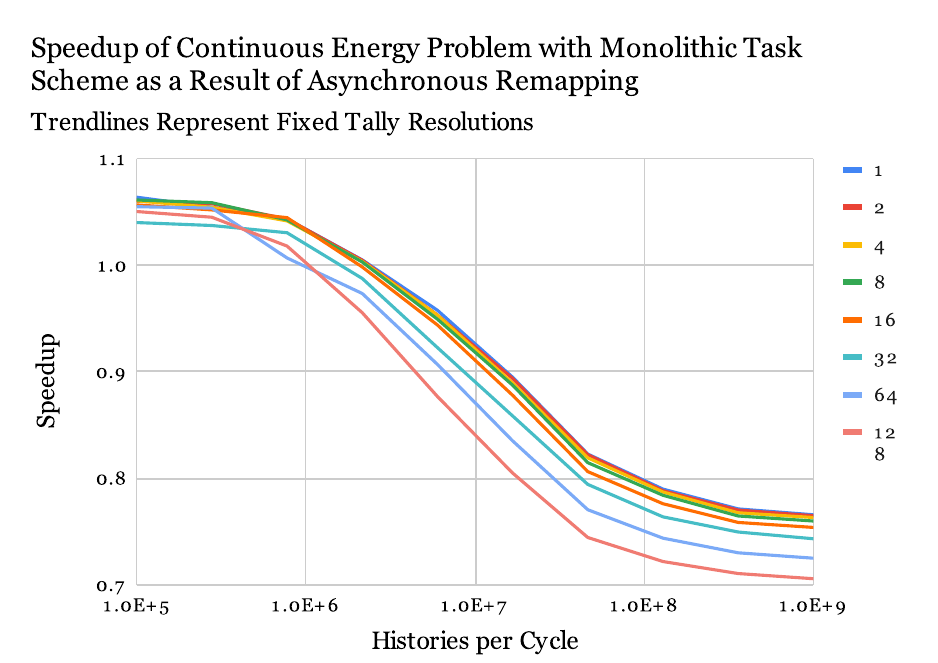}
    \end{minipage}
    \caption{Graphs depicting the speedup conferred by the decomposed task scheme with stack-based remapping (left) and the speedup conferred by asynchronous remapping with the monolithic task scheme (right) relative to the monolithic task scheme with stack-based remapping.}
    \label{fig:async_speedup}
    \label{fig:decomp_speedup}
\end{figure}

Asynchronous remapping exhibits reduced performance in the majority of the surveyed problem space under the monolithic task scheme, with speedup decreasing both as the particle count and tally resolution increase (Fig \ref{fig:async_speedup}).
We propose the inverse correlation between tally resolution and performance is likely due to the lower warp occupancy of streaming multiprocessors (SMs) under asynchronous mapping.
With the use of stashes, asynchronous mapping has a higher per-warp footprint in shared memory, reducing the total number of warps that can be concurrently executed.
Higher resolution tally meshes likely reduce the number of warps eligible for execution by requiring additional atomic operations, which in turn require the executing warp to wait for those operations.
This would have a greater impact on asynchronous remapping, as its reduced warp occupancy lowers the number of idle warps required to eliminate all warps from eligibility.

The more significant impact of particle count indicates less competitive performance in the most common paths of execution, and hence in the remapping itself. Given that the asynchronous remapping scheme provides better performance at smaller particle counts, this is unlikely to be the result of a performance penalty incurred by the scheme itself. As noted in \cite{pozulp_progress_2023}, larger particle counts remapped with larger banks/stacks can amortize the fixed costs of stack-based remapping across a larger amount of data, deriving better efficiency with larger remapping operations. Hence, the amortization of stack-based remapping at higher particle counts naturally results in a drop in relative performance for asynchronous remapping.

As with the use of asynchronous remapping on a the monolithic task scheme, the use of stack-based remapping on the decomposed task scheme results in reduced performance as the particle count increases (Fig \ref{fig:decomp_speedup}).
Unlike the previous comparison, there is no initial speedup above one, indicating an overall decrease in remapping performance.
This makes sense, given the amortization of remapping costs with larger data sets.
While the number of particles remains the same, each event type must track its own stack of particles, and the increased number of stacks divides the available device memory across more buffers, reducing per-stack capacity.
This explanation is further supported by the inflection point between \num{1e6} and \num{1e7}particle histories in fig \ref{fig:decomp_speedup}, where speedups begin to decrease more dramatically.
Between these two counts is each stack's provisioned capacity under the decomposed task scheme, \num{2e6}.

Unlike the particle count, the tally resolution correlates positively with the conferred speedup under the decomposed task scheme, and this correlation grows with the particle count. 
This aligns with expectations, since the overhead of accumulating these tallies would begin to dominate as resolution increases, marginalizing the performance gap in the rest of the simulation.

\begin{figure}
    \centering
    \includegraphics[width=\textwidth]{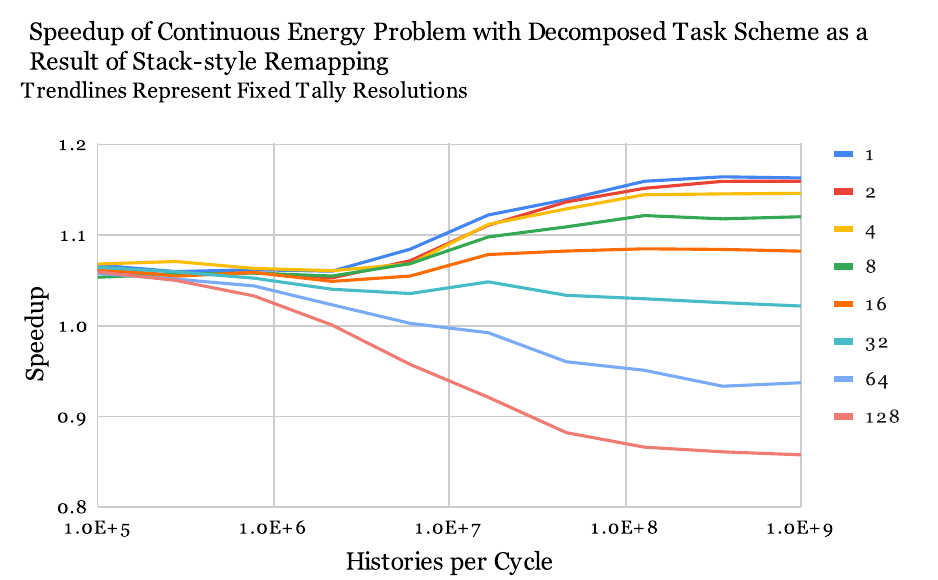}
    \caption{A graph depicting the speedup conferred by the joint use of the decomposed task scheme and asynchronous mapping relative to the monolithic task scheme with stack-based remapping.}
    \label{fig:combo_speedup}
\end{figure}

Interestingly, while the use of asynchronous remapping and the decomposed task scheme individually result in poor performance, their use in combination provides superior performance in some cases (Fig \ref{fig:combo_speedup}).
This counter-intuitive speedup is due to asynchronous remapping \textit{not} deriving much of its performance from amortizing costs over larger particle counts.
Since asynchronous remapping is reasonably cheap at most particle counts, it can much more easily benefit from the divergence reduction of further decomposition.

Of course, the relative performance of this scheme still drops as tally resolution increases, which is consistent with our warp occupancy theory.
However, this performance boost from decomposition offers an avenue for research.
Alternate decompositions of the transport loop could provide even better speedup.
Additionally, the warp occupancy issue could be resolved through more advanced data structures.
With the introduction of lock-free stash transactions, the same stash may be used by multiple warps, allowing for higher occupancy.

\section{Conclusions} \label{sec:6}

Both of the remapping algorithms available on MC/DC offer competitive performance relative to each-other for problem types that suit their strengths.
By providing cheap remapping at smaller history counts, asynchronous remapping provides strong scaling well suited for problems that are run at a smaller scale, setups that spread particles more thinly across hardware, and event systems that more finely subdivide transport logic.
In contrast, stack-based remapping offers relatively low per-warp overhead and economy at scale, capable of matching or exceeding the performance of asynchronous remapping in many cases as particle counts increase. 

\section*{Acknowledgements}
This work was supported by the Center for Exascale Monte-Carlo Neutron Transport (CEMeNT) a PSAAP-III project funded by the Department of Energy, grant number: DE-NA003967.

The authors would like to thank the Livermore computing staff for continued support using Lassen.
%The authors would like to thank Dr. Steven Hamilton for supplying the C5G8 cross section data as well as the Livermore computing staff for continued support using Dane, Tioga, and Lassen.
%The authors would also like to thank Damon McDougall and Dominic Encantre from AMD for support with Numba HIP and AMD compilers.

%\nocite{*}

\bibliographystyle{IEEEtran}
\bibliography{refs}
\newpage

\end{document}